\documentclass[showpacs,prl,twocolumn,aps,superscriptaddress]{revtex4}
 \usepackage{bm} \usepackage{amsmath} \usepackage{amssymb} \usepackage{latexsym}
 \usepackage{amsfonts} \usepackage{graphicx}

   \newcommand\romand{\mathrm{d}}

      \newcommand{\bea}{\begin{eqnarray}}
      \newcommand{\ea}{\end{eqnarray}} \newcommand{\eea}{\end{eqnarray}}

\begin{document}
\title{Atomic Bloch-Zener oscillations for sensitive force measurements in a cavity}
\author{B.\ Prasanna Venkatesh}
\affiliation{Department of Physics and Astronomy, McMaster University, 1280 Main St.\ W., Hamilton, ON, L8S 4M1, Canada} \author{M.\ Trupke}
\affiliation{Centre for Cold Matter, Imperial College, Prince Consort Road, London, SW7 2AZ, United Kingdom}
\author{E.\ A.\ Hinds}
\affiliation{Centre for Cold Matter, Imperial College, Prince Consort Road, London, SW7 2AZ, United Kingdom}
\author{D.\ H.\ J.\ O'Dell}
\affiliation{Department of Physics and Astronomy, McMaster University, 1280 Main St.\ W., Hamilton, ON, L8S 4M1, Canada}

\begin{abstract} Cold atoms in an optical lattice execute Bloch-Zener oscillations when they are accelerated.
We have performed a theoretical investigation into the case when the optical lattice is
the intra-cavity field of a driven Fabry-Perot resonator.
When the atoms oscillate inside the resonator, we find that their back-action
modulates the phase and intensity of the light transmitted through the cavity.
We solve the coupled atom-light equations self-consistently and show that,
remarkably, the Bloch period is unaffected by this back-action.
The transmitted light provides a way to observe the oscillation continuously,
allowing high precision measurements to be made with a small cloud of atoms. \end{abstract}

\maketitle
When quantum particles in a potential lattice are subjected to a constant force $F$, they execute
Bloch-Zener oscillations (BZOs) \cite{bloch28} with a frequency
\begin{equation}
\omega_{\mathrm{B}}=Fd/\hbar\,, \label{eq:BZOfreq}
\end{equation}
where $d$ is the period of the lattice.
This behavior was first demonstrated \cite{superlattice} with electrons in semiconductor superlattices, where a DC electric field
provided the force. However, rapid dephasing due to impurities \cite{reynolds96} has  prevented BZOs from becoming useful in solid state devices.

Cold atoms in optical lattices have recently provided an alternative realization of BZOs
\cite{salomon,raizen,anderson98,morsch01,roati04,battesti04,ferrari06} in which long
coherence times are possible.
Initially, acceleration of the lattice induced the oscillations, but in subsequent experiments
\cite{anderson98,roati04,ferrari06}, gravity provided the required force.
In \cite{ferrari06}, the BZO damping time was 12s, allowing some 4000 cycles to be measured
over 7s. With such long coherence times, cold atom BZOs become suitable for high precision
measurements, for example to determine the fine structure constant $\alpha$ \cite{battesti04},
to measure gravity \cite{ferrari06}, or to explore Casimir-Polder forces \cite{carusotto05}.
In the experiments to date, it has been necessary to reconstruct the oscillations by making
destructive measurements at a large number of different times, each measurement requiring
a new cloud of atoms to be trapped, cooled, and loaded into the lattice.
The process is laborious and suffers from shot-to-shot variations in the initial cloud conditions.
In this letter, we discuss how the measurement could be substantially improved by using an optical cavity to enhance the interaction of the atoms with the light.
We show how the light transmitted through the cavity can provide an \textit{in vivo}
observation of the BZOs and we assess the extent to which this
perturbs the motion of the atoms. Finally we consider the statistical sensitivity of the
method and show that it can yield high precision in a single shot.

\begin{figure}
\includegraphics[width=0.4\textwidth]{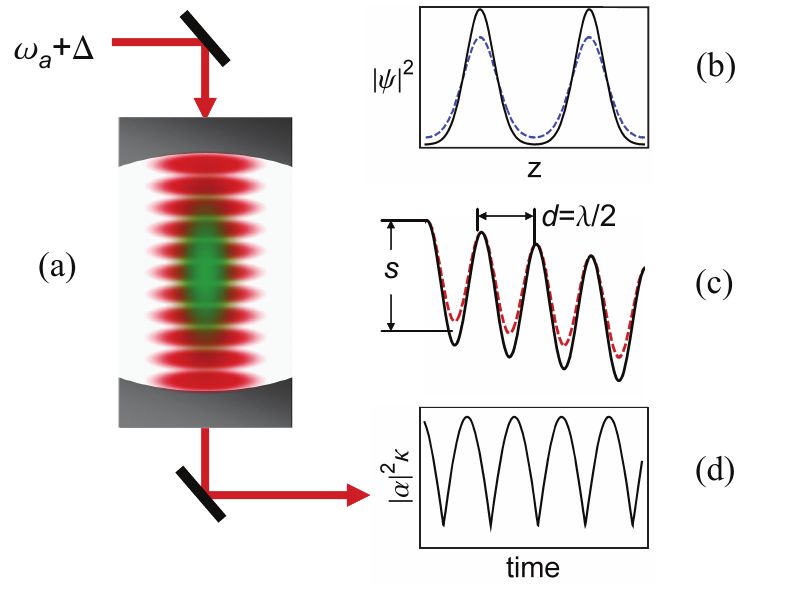}
\vspace{-0.5cm}
\caption{Schematic of the proposed experiment.  (a) A cloud of cold atoms is held in a standing-wave optical trap inside a vertical Fabry-Perot cavity.   (b) The atoms execute Bloch-Zener oscillations, leading to a periodic modification of their wave function. (c) This modulates the intra-cavity power and hence the lattice depth $s$. (d) The power modulation is seen in the light transmitted by the cavity.} \label{fig:schematic}
\end{figure}

Let us take the cavity to be a vertical Fabry-Perot resonator illuminated by a laser, which
makes a standing-wave light field inside, shown in Fig.\,\ref{fig:schematic}(a).
Cold atoms in this lattice execute BZOs under the influence of gravity, as illustrated in Fig.\,\ref{fig:schematic}(b).
The optical dipole interaction between one cavity photon and an atom placed at an
antinode of the field is given by
$\hbar g_{0}=\mu \sqrt{\hbar \omega_{\mathrm{c}}/(\epsilon_{0}V)}$,
where $\omega_c$ and $V$ are the frequency and volume of the relevant cavity mode
and $\mu$ is the atomic transition dipole moment.
The effect of one atom on the cavity field is characterised by the
cooperativity $C=g_{0}^2/(2\kappa \gamma)$, where $2\gamma$ is the atomic spontaneous
emission rate in free space and $2\kappa$ is the cavity energy damping rate. When $C\gtrsim 1$, the cavity field is strongly perturbed by the atom, as illustrated in Fig.\,\ref{fig:schematic}(c).
Thus the light transmitted or reflected by such a cavity can detect the presence of a
single atom \cite{hood98}, and can be sensitive
to the motion of atoms trapped within the cavity \cite{ye99}, as in Fig.\,\ref{fig:schematic}(d). Although we shall not discuss Bose-Einstein condensates (BECs) in this paper, we note in passing that several experiments have succeeded in placing BECs inside optical cavities \cite{cavitybec}.

Consider a cavity mode, whose frequency in the absence of any atoms is $\omega_c$,
 pumped by an external laser of the same frequency. A cloud of $N$ atoms, each of mass $m$, is placed inside the cavity and is sufficiently dilute that the atoms do not interact directly with each other.
 The coupled atom-cavity hamiltonian becomes \cite{maschler05}
\begin{eqnarray}
\hat{H} & = &\frac{\hbar^2}{2m}\int \vert \nabla_{z} \hat{\psi}\vert^{2}  \romand z +
\frac{\hbar g_{0}^{2}}{\Delta}\hat{a}^{\dag}\hat{a}\int \hat{\psi}^{\dag} \hat{\psi} \,  \cos^{2}(k_{\mathrm{c}}z) \romand z \nonumber \\
&& +F\int  \hat{\psi}^{\dag} \hat{\psi} \, z \ \romand z
-\mathrm{i} \hbar (\eta^{\ast}\hat{a}-\eta \hat{a}^{\dag})
\, \label{eq:hamiltonian}
 \end{eqnarray}
where the operator $\hat{a}(t)$ annihilates a photon in the cavity mode and the operator $\hat{\psi}(z,t)$ annihilates an atom at the point $z$.
The first term gives the kinetic energy of the atom. The second describes the
quadratic Stark interaction in the rotating-wave approximation.
Here, $\Delta=\omega_{\mathrm{c}}-\omega_{a}$ is the detuning
between the cavity mode and the atomic transition frequency $\omega_{a}$, and $k_c=\omega_c /c$.
This is an approximate form, that is valid when $\Delta\gg\gamma$
and $\eta^2\ll(\kappa\Delta/g_0)^2$. Under these conditions the atom has negligible
population in the excited state. The third term accounts for the external force, and the
last term describes the coherent excitation of the cavity
 by the external laser. For a cavity with equal mirror reflectivities, the pumping rate is $\eta=\sqrt{\kappa I}$,
where $I$ is the rate of incident photons matching the cavity mode. The cavity field is a driven and damped quantum harmonic oscillator for which it is known that exact solutions of the Fokker-Planck equation are coherent states \cite{louisell67}.
Furthermore, since there is negligible spontaneous emission by the atoms, the cavity field remains in a coherent state in the presence of atoms. Taking the expectation value of the Heisenberg equations of motion for $\hat{a}$ and $\hat{\psi}$ in the coherent state $\vert \alpha \rangle$ yields the
equations of motion \cite{horak01}:
\begin{eqnarray}
 \dot{\alpha} & = &-\mathrm{i} \frac{\alpha N g_{0}^{2}}{\Delta}\int \vert \Psi \vert^{2} \cos^{2}(k_{\mathrm{c}}z) \romand z
 + \eta-\kappa\alpha\,,
\label{eq:eqnofmotionlight} \\ \mathrm{i}\hbar \dot{ \Psi} & = & \left(\frac{-\hbar^{2}}{2m}\partial_{z}^{2}
 +\frac{\hbar g_{0}^2 \vert \alpha \vert^{2}}{\Delta}\cos^{2}(k_{\mathrm{c}}z) +Fz \right)\Psi
 \label{eq:eqnofmotionatom}
 \end{eqnarray}
where $\Psi(z,t)=\langle \Psi_{N-1} \vert \hat{\psi}(z,t) \vert \Psi_{N} \rangle/\sqrt{N}$ is the wave function occupied by all $N$ atoms and $\vert \Psi_N \rangle$ is the corresponding state vector in Fock space. We have added a damping term proportional to $\kappa$ in Eq.\ (\ref{eq:eqnofmotionlight}) to account for leakage of light through the mirrors \cite{cct}. These equations neglect quantum fluctuations of the light field which can heat the atoms: we return to this effect later.
Equations\,(\ref{eq:eqnofmotionlight}) and (\ref{eq:eqnofmotionatom}) must be solved self-consistently: the coupling
\begin{equation}
g^{2}(t)= g_{0}^{2} \int \vert \Psi (z,t) \vert^{2} \cos^{2}(k_{\mathrm{c}} z) \romand z \label{eq:coupling}
\end{equation}
changes $\alpha$, which changes the depth of the lattice.
This alters the atomic wave function and therefore changes $g$, etc. In static equilibrium, $\dot{\alpha}=0$ and then
\begin{eqnarray}
\alpha=\frac{\eta}{\kappa}\,\,\,\frac{1}{1+\mathrm{i} N g^{2}(t)/(\kappa\Delta)} . \label{eq:steadystate}
\end{eqnarray}
Even if the atoms are in motion, (\ref{eq:steadystate}) remains a very good approximation
since $\kappa$ is generally much greater than the highest frequency in the atom dynamics,
so that the field `instantaneously' adapts to the atomic distribution. Frequencies that feature in the atomic motion are the BZO frequency, the band splitting, and  the harmonic frequency $\omega_{\mathrm{ho}}=2 g_{0} \vert \alpha \vert \sqrt{E_{\mathrm{R}}/(\hbar \Delta)}$ at the bottom of each potential well, $E_{\mathrm{R}}=\hbar^{2} k_{\mathrm{c}}^{2}/(2m)$ being the atomic recoil energy.
For the experiments we consider here $\omega_{\mathrm{B}} $ and $ \omega_{\mathrm{ho}}$ are much smaller than $\kappa$, so we assume in our analytic calculations, though not in our numerical simulations, that Eq.\,(\ref{eq:steadystate}) holds.

Let us recall the standard theory of BZOs without a cavity  \cite{holthaus96,gluck98,holthaus00,zapata01,thommen02,hartmann04,larson06}.
The Schr\"{o}dinger Eq.\ (\ref{eq:eqnofmotionatom}) can be written
as $\mathrm{i}\hbar \dot{\Psi}=(\hat{H}_{\mathrm{0}}+Fz) \Psi$, where $\hat{H}_{0}=\hat{p}^{2}/(2m)
+s \cos^{2}k_{\mathrm{c}}z $.
The eigenfunctions $ \chi_{q,s,n}(z)$ of $\hat{H}_{0}$ are Mathieu functions (Bloch waves) that in general depend on position $z$,
lattice depth $s$, band index $n$ and quasimomentum $q$, restricted to the first Brillouin zone $-\pi/d \leq q \leq \pi/d$ \cite{a+s}. We assume in our analytic calculations that the atoms are in the lowest band,
whose energy is $E_{q,s}$, and dispense with the band index so that $ \hat{H}_{0} \chi_{q,s}=E_{q,s}\chi_{q,s}$.
The Bloch theorem allows us to write $\chi_{q,s}(z)=U_{q,s}(z)\exp[\mathrm{i}qz]$,
where $U_{q,s}(z)=U_{q,s}(z+d)$ obeys
\begin{equation} \frac{(\hat{p}+\hbar q)^{2}}{2m}U_{q,s}(z)
+s \cos^{2}(k_{\mathrm{c}}z) U_{q,s}(z)=E_{q,s} U_{q,s}(z). \label{eq:BlochEq} \end{equation}
To tackle the full hamiltonian $\hat{H}_{0}+Fz$ we make the gauge
 transformation $\Psi(z,t)=\exp[-\mathrm{i}Ft z/\hbar]\tilde{\Psi}(z,t)$, yielding
the Schr\"{o}dinger equation  $\mathrm{i}\hbar\dot{\tilde{\Psi}}
=\tilde{\hat{H}}\tilde{\Psi}$, where $\tilde{\hat{H}}=\mathrm(\hat{p}-Ft)^{2}/(2m)+s \cos^{2} k_{c}z$.
Comparing this with (\ref{eq:BlochEq}) we see that the effect of the force is to evolve the quasi-momentum  according to  Bloch's acceleration theorem \cite{bloch28}
\begin{equation} q \rightarrow q(t)=q_{0}-Ft/\hbar\,, \label{eq:BlochAcclThm}
\end{equation}
where $q_{0}$ is the quasimomentum at $t=0$.
 When $q(t)$ reaches the edge of the
Brillouin zone at $-\pi/d$ it is
mapped to the identical point $q=+\pi/d$, giving rise to oscillatory behaviour - the BZO.
The corresponding Bloch wave
has the approximate form \cite{houston40} (setting $q_0 = 0$)
\begin{equation} \tilde{\Psi}(z,t) \approx U_{q(t),s}(z) \exp[-\mathrm{i}/\hbar \int^{t} \romand t' \ E_{q(t'),s}]\,,
\label{eq:BlochWave}
\end{equation}
within the adiabatic approximation that the rate of change $\dot{U}/U$ is too small to excite higher bands.
Here $U_{q(t),s}(z)$ is the instantaneous solution of Eq.\ (\ref{eq:BlochEq}).
During a BZO the spatial distribution $U_{q(t),s}(z)$  oscillates with a breathing motion, as shown schematically in Fig.\ 1(b).

Consider now the effect of the BZOs on the field inside the cavity.
The coupling $g$ (Eq.\,(\ref{eq:coupling})) depends on $|\Psi(z,t)|^2$ which equals $|\tilde{\Psi}(z,t)|^2$.
Its breathing motion changes $g$, which in turn modulates the cavity field through Eq.\,(\ref{eq:steadystate}).
Inserting (\ref{eq:steadystate}) into (\ref{eq:eqnofmotionatom}), and
replacing $\eta$ by $ \sqrt{\kappa I}$, we obtain the
Schr\"{o}dinger equation for atoms in a periodic potential $\mathrm{i}\hbar \dot{ \Psi} =
(\hat{p}^{2}/2m+s(t)\cos^{2}(k_{\mathrm{c}}z) +Fz)\Psi $, with the time-dependent potential depth
\begin{equation} s(t)=
\hbar I \left(\frac{g_{0}^{2}}{\Delta  \kappa}\right)\frac{1}{1+ \left(N g^{2}(t)/(\Delta  \kappa)\right)^{2}}.
\label{eq:latticeamplitude}
\end{equation}

\begin{figure}[htb!]
\includegraphics[width=0.4\textwidth]{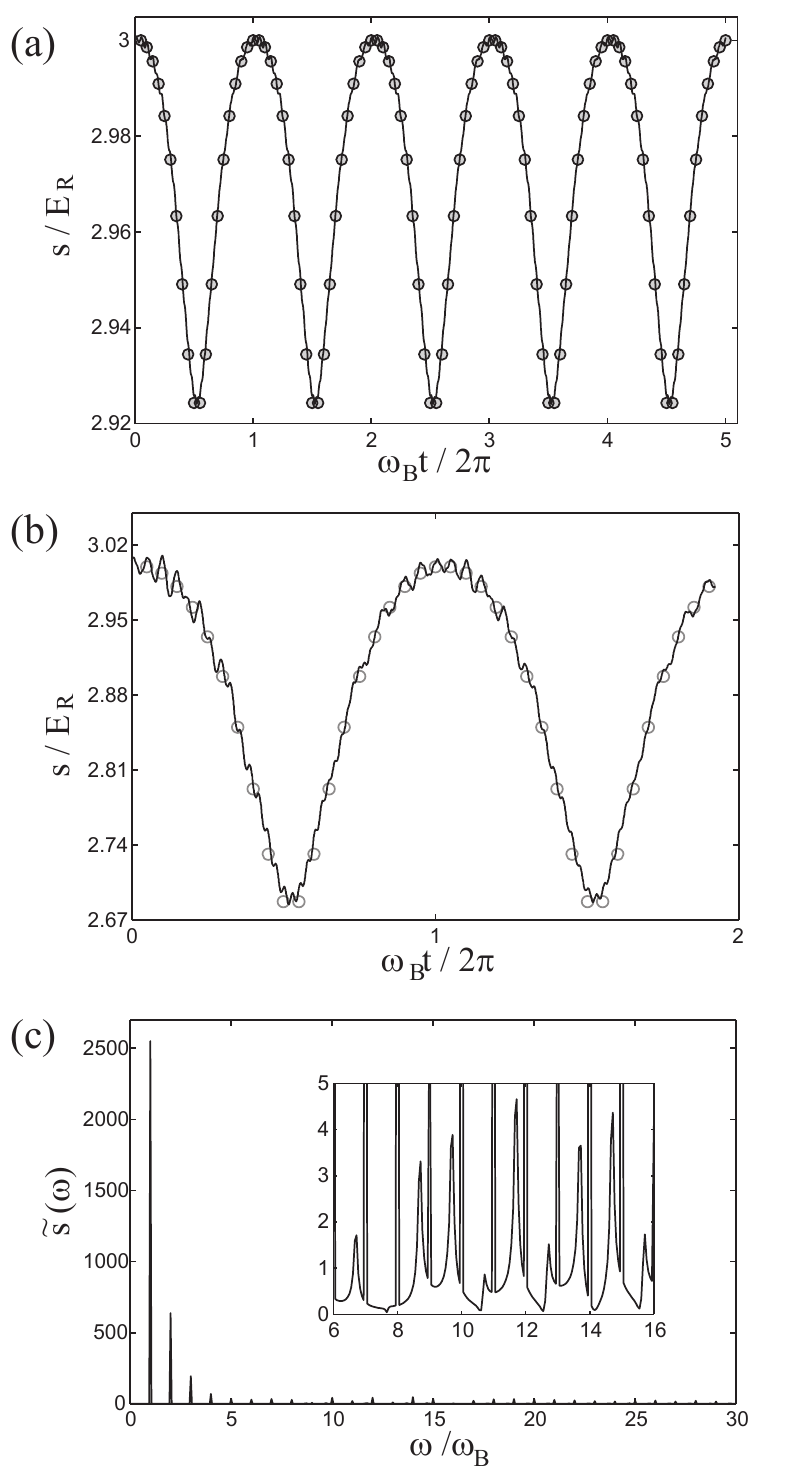}
\vspace{-0.5cm}
\caption{Calculated evolution of lattice depth $s(t)$ normalized to the recoil energy for $^{87}$Rb atoms undergoing 840 Hz Bloch-Zener oscillations in a 780\,nm lattice with $s\approx 3 E_R$. Lines: numerical solution to Eqs.\,(\ref{eq:eqnofmotionlight}) and (\ref{eq:eqnofmotionatom}). Dots: self-consistent adiabatic approximation of Eqs.\,(\ref{eq:BlochWave}) and  (\ref{eq:latticeamplitude}). (a) $N g_{0}^{2}/(\kappa \Delta)=0.4$. (b) Close-up of lattice depth oscillations with stronger coupling, $N g_{0}^{2}/(\kappa \Delta)=1$. Non-adiabatic effects are seen in the line. (c) Fourier transform $\tilde{s}(\omega)$ of result in (b), showing harmonics of fundamental frequency $\omega_{\mathrm{B}}$. Inset: Close-up of $\tilde{s}(\omega)$ at higher frequencies. Harmonics of $\omega_{\mathrm{B}}$ appear as sharp vertical lines. In addition, one sees much weaker, broad, Fourier components due to band excitation, corresponding to the rapid oscillations in (b). The actual individual values of the parameters used (or assumed in the case where only ratios enter) in the calculations were (see text): $N=5 \times 10^{4}$, $g_{0}=2 \pi \times 2.8$ MHz, $\kappa= 2 \pi \times 1.0$ MHz. In (a) $\Delta=2 \pi \times 1.0$ THz, $\eta= 2 \pi \times 39 $ MHz; in (b) and (c) $\Delta=2 \pi \times 0.39$ THz, $\eta= 2 \pi \times 28 $ MHz.} \label{fig:oscillation}
\end{figure}

When $N g_{0}^{2}/(\kappa\Delta)\ll1$, $s$ is approximately constant in time and
$\alpha$ becomes $(\eta/\kappa)[1-i N g(t)^2/(\Delta\kappa)]$, i.e. the light exhibits
a small phase modulation and has negligible intensity variation.
In this case, the atoms oscillate in a lattice that is essentially static.
With stronger coupling, where $N g_{0}^{2}/(\kappa\Delta)\simeq 1$, $s(t)$ is changed 
significantly during an oscillation. The fundamental period is nevertheless unchanged and is still given by Eq.\ (\ref{eq:BZOfreq}). Physically, this is because BZOs arise from an interference of waves in a lattice akin to Bragg scattering and lattice depth plays no role in determining the phase-matching condition. Rather, this is determined by the symmetry of the hamiltonian which is precisely maintained at all times. To examine the effect of lattice depth modulation in more detail, let us begin with the adiabatic case where the frequency spectrum of $s(t)$ remains largely
at low frequencies unable to excite higher Bloch bands. An example is shown in Fig.\ \ref{fig:oscillation}(a).
In that case, the lowest band energy $E_{q,s(t)}$ is still determined by
the instantaneous value of $s(t)$, with $H_{0}(t)\chi_{q,s(t)}=E_{q,s(t)}\chi_{q,s(t)}$,
and $q_{0}$ remains a constant of the motion generated by $H_{0}(t)$
despite the time-dependent potential (\cite{holthaus96} reaches a similar
conclusion for electrons in an ac field). Consequently, the Bloch wave in the
presence of an external force can still be calculated using Eq.~(\ref{eq:BlochWave}),
provided we use the instantaneous values of $s(t)$ and $s(t^{\prime})$.
It only remains to find the self-consistent solution for $s(t)$ by solving
Eq.\ (\ref{eq:latticeamplitude}) at each instant of time. Here $s(t)$ appears
explicitly on the left and also implicitly on the right as a parameter determining
the Bloch wave function that is required to calculate the coupling
$\sqrt{N}g(t)$ using Eq.\ (\ref{eq:coupling}). We conclude that despite
the lattice depth modulation, the Bloch acceleration
theorem (\ref{eq:BlochAcclThm}) still holds for atoms in the lowest band and therefore the fundamental
oscillation frequency remains identical to the $\omega_{\mathrm{B}}$ of an
atom in a static lattice. Furthermore, because $\omega_{\mathrm{B}}$ is the same for all
bands, there is no frequency shift even when higher bands are excited, as we
have verified numerically for a wide range of $N g_{0}^{2}/(\Delta \kappa)$.

Fig.\ \ref{fig:oscillation}(a) compares the lattice depths obtained by numerical solution
of Eqs.\,(\ref{eq:eqnofmotionlight}) and (\ref{eq:eqnofmotionatom}) (lines)
and by the adiabatic approximation of Eqs. (\ref{eq:BlochWave}) and
(\ref{eq:latticeamplitude}) (dots). The two are in good agreement. Figure\,\ref{fig:oscillation}(b), shows a case of stronger coupling, where one can see fast oscillations superimposed on the main motion. The Fourier transform
of this reveals two effects, illustrated in Fig.\,\ref{fig:oscillation}(c).  (i) There are higher
harmonics of $\omega_{\mathrm{B}}$ because the oscillations at the Bloch periodicity
are not exactly sinusoidal. In the adiabatic solution, the first four are accurately
reproduced and the higher harmonics are very small. (ii) The exact solution of
Eqs.\,(\ref{eq:eqnofmotionlight}) and (\ref{eq:eqnofmotionatom}) predicts
non-adiabatic components at higher frequencies, shown inset in Fig.\,\ref{fig:oscillation}(c). These are predominantly harmonics
of $\omega_{\mathrm{B}}$ (the sharp lines), but in addition, there are other frequency components that can be seen as broad lines. These are only found in the full numerical solution and are due to a small amount of excitation to higher Bloch bands. These non-adiabatic effects become stronger as the parameter $N g_{0}^{2}/(\kappa \Delta)$ is increased.

The atoms are driven not only by the mean-field potential $s(t) \cos^{2}(k_{c}z)$, but also by random forces due to i) spontaneous emission, and ii) fluctuations in the photon number, that are associated with the decay rates $\gamma$ and $\kappa$, respectively. In particular, in the strong coupling regime photon number fluctuations can significantly heat atoms inside an optical cavity \cite{ye99}.  We quantify the heating effect via the increase in the width $\sigma_{p}$ of the atom's momentum distribution according to $\mathrm{d} \,  \sigma_{p}^{2} / \, \mathrm{d} \, t=2D$. The diffusion constant $D= \int_{0}^{\infty} dt' [\langle F_{\mathrm{dip}}(t)F_{\mathrm{dip}}(t+t')\rangle-\langle F_{\mathrm{dip}} \rangle^2] $  \cite{gordon&ashkin80}  involves two-time correlations of the dipole force $F_{\mathrm{dip}}$, and hence of the intracavity electric field. It has been calculated for atoms in cavities in \cite{hechen98} and has two terms. The first occurs in any standing-wave light field and at low saturation is given by $D_{\mathrm{sw}}=\hbar^{2} k^{2}/ 2 \tau_{\mathrm{sp}}$ \cite{gordon&ashkin80}, where $\tau_{\mathrm{sp}}^{-1} = 2  \gamma |\alpha|^2 g_{0}^{2}/\Delta^2$ is the spontaneous emission rate at an antinode. The second term, specific to cavities, is $D_{\mathrm{cav}}=2 D_{\mathrm{sw}} \,C \sin^{2}(2 k_{c}z)$. This diffusion limits the coherent measurement time, which we take to be the time $\tau$ when the momentum distribution has a width equal to one half of the first Brillouin zone, i.e. $\sigma_{p}=\hbar k_{c}$. Then $\tau= \tau_{\mathrm{sp}}/(1+C)$, where we have replaced  $\sin^{2}(2k_{c}z)$ by $1/2$ - a good approximation in the ground band for a lattice of depth $s=3 E_{\mathrm{R}}$.

The BZOs can be observed by detecting the photon current $|\alpha(t)|^2\kappa$ transmitted
through the cavity which is directly proportional to the depth $s(t)$ of the lattice ($s= \hbar g_{0}^{2}|\alpha|^{2}/\Delta$, see Eq.\ (\ref{eq:eqnofmotionatom})), whose evolution is shown in Fig.\ \ref{fig:oscillation}. 
 For an estimate of the measurement precision, let us write the detection
rate as $R[1+\epsilon\, \cos(\omega t)]$. After measuring this for a time $\tau$ with detectors having an efficiency $\xi$, the
shot noise gives an uncertainty in the oscillation frequency of
$\sigma_{\omega}\approx 2\pi \tau^{-3/2}/(\epsilon\sqrt{\xi R})$, in which $\tau^{-1}$ comes
from the linewidth due to the finite duration of the measurement, and $\tau^{-1/2}$ comes
from the shot noise in this bandwidth. This simple estimate is close to the Cram\'{e}r-Rao lower bound~\cite{Kay},
the limit given by the information content of the signal. For
small $\epsilon$, $R\approx|\alpha|^2\kappa$, then the frequency uncertainty
 can be written as $\sigma_\omega\approx2\pi\frac{s}{\hbar}\frac{1}{\epsilon \sqrt{\xi}}(\frac{g_{0}^2}{\kappa \Delta})^{2}(\frac{1}{C}+1)^{3/2}$. In order to bring out the implicit dependence on the number of atoms $N$ contained in this result, we define the parameter $x=N g_0^2/(\kappa\Delta)$. Referring to Eq.\ (\ref{eq:latticeamplitude}), 
if the number of atoms 
is increased then proportional increases in the laser detuning and intensity
maintain constant values of $s$, $x$ and $\epsilon$, while the measurement time
$\tau$ and the intra-cavity power both increase by the factor $N$. With this scaling,
the measurement time, detuning and frequency uncertainty are given by
\begin{eqnarray}
\tau & = & \frac{\hbar}{s x}\frac{NC}{1+C} \label{eq:measurementtime} \\
\Delta & = & \frac{2\gamma}{x} N C\\
\sigma_\omega & \approx & \frac{2\pi s x^2} {\sqrt{\xi}\hbar\epsilon}\frac{1}{N^2} \left(\frac{1}{C}+1\right)^{3/2}. \label{eq:sigma}
\end{eqnarray}
The uncertainty $\sigma_\omega$ therefore decreases rapidly with a dramatic $1/N^2$ dependence which ultimately derives from the continuous observation of the oscillations (via the enhanced measurement time $\tau$).

Small $s$ reduces $\sigma_\omega$, but the lattice must support the atoms against gravity. We find that $s=3E_R$ is a reasonable compromise. In the example of Fig.\,\ref{fig:oscillation}(a) we have chosen $x=0.4$, which gives $\epsilon =1.3\%$ in this lattice. Taking a reasonable number of atoms, let us say $5\times 10^4$, a readily achieved cooperativity of $C=1.3$, and a photon detector with $60\%$ efficiency, brings $\sigma_\omega/\omega_B$ to 1ppm. 
From the definitions of $x$ and $C$ this requires a detuning of $\Delta=2\pi\times1\,$ THz and from Eq.\ (\ref{eq:measurementtime}) a measurement time  of only $\tau=1$\,s. 
These numbers fix the ratio $g_{0}^2/ \kappa = 2\pi\times 7.8\,$ MHz. If we choose $\kappa= 2 \pi \times 1$ MHz, then using $s= \hbar g_{0}^{2}|\alpha|^{2}/\Delta$, this means there are on average $> 1400$ photons in the cavity.

We have not included direct atom-atom interactions in the model discussed here. These can lead to quasimomentum-changing transitions (non-vertical transitions in the language of \cite{mueller02}) which dephase the BZOs. However, as summarized in the opening paragraphs of this paper, long-lived BZOs have already been successfully demonstrated in cold gases containing many atoms, and so it is a question of degree, i.e.\ at what atom density and interaction strengths do the interactions become important? An experiment investigating the control of interaction-induced dephasing of BZOs in a Bose-Einstein condensate has recently been reported \cite{Gustavsson08}. Using a Feshbach resonance they were able to increase their dephasing time from a few to more than 20 thousand BZO periods. 
From the details of their measurements we estimate that, for our example given immediately above involving $5 \times 10^{4}$ atoms, the dephasing due to collisions is negligible for reasonable cavity geometries. The effect becomes significant on increasing the number of atoms to several million, but can be suppressed by tuning to a Feshbach resonance \cite{Gustavsson08}. Large detuning and laser power impose a practical limit on the useful atom number at about this level anyway.

For atoms being continuously measured, an important source of dephasing is quantum measurement back-action. This effect is included in the estimate above in a quasi-classical way through the diffusive heating of the atoms by fluctuations of the cavity light field. The cavity field suffers fluctuations because it is dissipatively coupled to the outside world and it is precisely the light escaping from the cavity (at rate $\kappa$) that contains the information about the state of the atoms. In other words, it is the cavity decay that is doing the measuring. For our parameters, $\tau < \frac{1}{2}\tau_{\mathrm{sp}}$ because $C >1$, and therefore it is the fluctuations due to cavity decay that limit the measurement time, rather than the spontaneous emission.
We plan to perform a more microscopic study of the measurement back-action in the future. 

In conclusion, we predict that the force on a small cold atom cloud can be measured very accurately by a new method based on BZO oscillations in an optical cavity. The BZO oscillations are measured continuously by monitoring the light that leaks out of the cavity. This enables a relatively fast and high precision measurement of the oscillation frequency,
the error being given by Eq.\ (\ref{eq:sigma}).  Our treatment of the problem is based upon solving  the coupled equations of motion for the atoms and light. This gives a detailed picture of the dynamics, including the adiabatic and non-adiabatic aspects.

We note that since the submission of this paper a related proposal on monitoring of Bloch oscillations using an optical cavity has appeared \cite{peden09}.

We thank Pavel Abumov, Donald Sprung and Wytse van
Dijk for discussions. Funding was provided by NSERC (Canada), EPSRC, QIPIRC
and the Royal Society (UK). Part of this work was done during D.O.'s stay at the
  Institut Henri Poincare - Centre Emile Borel. He thanks this institution for hospitality
and support.

\end{document}